\documentclass[prl,aps,showpacs,twocolumn]{revtex4}
\usepackage{epsfig}
\usepackage{bm}

\begin{document}

\title{Magnetic field gradients in solar wind plasma and geophysics periods}

\author{A. Bershadskii}
\affiliation{ICAR, P.O. Box 31155, Jerusalem 91000, Israel}

\begin{abstract}
Using recent data obtained by Advanced Composition Explorer (ACE)
the pumping scale of the magnetic field gradients of the solar 
wind plasma has been calculated. This pumping scale is found to be equal to 
24h $\pm$ 2h. The ACE spacecraft orbits at the L1 libration point which 
is a point of Earth-Sun gravitational equilibrium about 1.5 million km 
from Earth. Since the Earth's magnetosphere extends into the vacuum of 
space from approximately 80 to 60,000 kilometers on the side toward the Sun 
the pumping scale cannot be a consequence of the 24h-period of the Earth's 
rotation. Vise versa, a speculation is suggested that for 
the very long time of the coexistence of Earth and of the solar wind the weak 
interaction between the solar wind and Earth could lead to stochastic 
synchronization between the Earth's rotation and the pumping scale of the 
solar wind magnetic field gradients. This synchronization 
could transform an original period of the Earth's rotation to the period close to the 
pumping scale of the solar wind magnetic field gradients.  
\end{abstract}


\maketitle

\section{Introduction}

Magnetic field in solar wind plasma is actively studied in the 
last years both theoretically and experimentally (see, for instance, 
\cite{barn1}-\cite{zmd}). Properties of this field gradients are 
of especial interest because of strong non-homogeneity of the field. 
Inferring universal properties of the magnetic field is a difficult 
task due to superposition of the strong non-homogeneity and global 
anisotropy. Even in the inertial range of scales the non-homogeneity 
and anisotropy affect behavior of different components of the magnetic 
field and its gradients. It can be shown \cite{bs}, however, 
that {\it magnitude} of the magnetic field $B = \sqrt{B_i^2}$ 
does exhibit certain universal properties in inertial range of scales. 
Moreover, we will show in present paper that magnitude of {\it gradients} of 
the solar wind magnetic field exhibits certain universal properties as well. 
These universal properties have substantial 
geophysical consequences, which we discuss in the last section of the paper.\\

In magnetohydrodynamics (MHD) the magnetic field fluctuation ${\bf
B}$ dynamics is described by equation
$$
\frac{\partial {\bf B}}{\partial t} = \nabla \times ({\bf v}
\times {\bf B}) + \eta \nabla^2 {\bf B}.    \eqno{(1)}
$$
Here, ${\bf v}$ is the turbulence velocity and the $\eta$ is
magnetic diffusivity. Equation (1) can be regarded as a vector
analogue of the advection-diffusion equation
$$
\frac{\partial \theta }{\partial t}=-({\bf v} \cdot \nabla) \theta
+ D \nabla^2 \theta              \eqno{(2)}
$$
for the evolution of a passive scalar $\theta$ subject to
molecular diffusivity $D$. Aside from the fact that ${\bf B}$ is a
vector and $\theta$ a scalar, the equations are different also
because $\bf v$ in Eq.\ (1) can be affected quite readily by the
feedback of the magnetic field $\bf B$. Our interest here is to
explore the extent of similarities, despite these obvious
differences, in the inertial (Batchelor) range statistics of the magnetic and
the passive scalar fields (see, for instance, Refs.\
\cite{shr}-\cite{ching} and the papers cited there).

\begin{figure} \vspace{-0.5cm}\centering
\epsfig{width=.45\textwidth,file=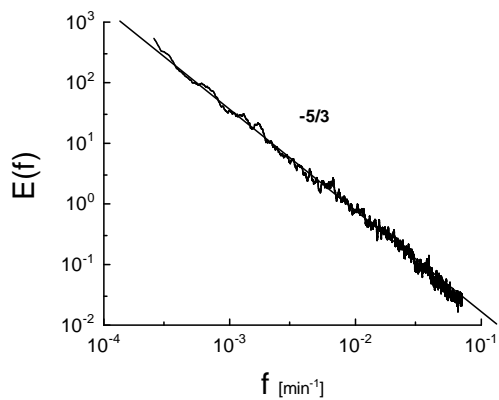} \vspace{-4cm}
\caption{Energy spectrum of the {\it magnitude} $B = \sqrt{B_i^2}$
of the magnetic field $\bf B$ in the solar wind plasma, as
measured by the ACE magnetometers in the nanoTesla range for the
year 1998 (4 min average).}
\end{figure}

Solar wind is an excellent natural ``laboratory" for the MHD
problem. It is known that the statistical properties of velocity
fluctuations in the solar wind are remarkably similar to those
observed in fluid turbulence \cite{bur}. It is also known that the
plasma power spectra of the magnetic field and velocity
fluctuations often contain an ``inertial'' range with a slope of
approximately $-5/3$ (for reviews, see \cite{bur},\cite{gold}).
The approximately $-5/3$ power-law is especially common for
magnitude fluctuations $B = \sqrt{B_i^2}$ (the summation over
repeated indexes is assumed) of the magnetic field, as one can see
in Fig.\ 1. For computing this spectrum, we have used the data
obtained from Advanced Composition Explorer (ACE) satellite
magnetometers for the year 1998. In this period, the sun was quiet
and the data are statistically stable. The range of scales for
which the ``$-5/3$" power holds is taken to be the inertial range;
the smaller scales are obliterated because of the instrument
resolution and the truncation at the large-scale end is governed
by the record length chosen for Fourier transforming.

The nature of the spectrum for each individual component of the
magnetic field is more variable from one component of $\bf B$ to
another, and from one situation to another, perhaps because of
large anisotropies in the magnetic field $\bf B$, but the result
for the $magnitude$ of $\bf B$ seems more robust. Scaling spectrum
with the $-5/3$ slope (Fig. 1) is quite typical of that observed
for passive scalar fluctuations in fully developed
three-dimensional fluid turbulence (the so-called Corrsin-Obukhov
spectrum \cite{my}). Spurred by this similarity, we were motivated
to explore further the properties of the magnitude $B$ and its gradients, 
and compare them with those of the passive scalar. 

\section{Statistical properties of the magnitude of the magnetic field}

More detailed statistical information is provided by the structure functions
scaling
$$
\langle |\Delta B_{\tau}|^p \rangle \sim \tau^{\zeta_p},
\eqno{(3)}
$$
where
$$
\Delta B_{\tau} =B(t+\tau)-B(t).       \eqno{(4)}
$$
The exponent $\zeta_2$ is directly related to the spectral
exponent (in our case $\zeta_2 \approx 5/3 - 1 = 2/3$ \cite{my}).
If the dependence of $\zeta_p$ on $p$ is nonlinear, it is
well-known that one has to deal with intermittency.

\begin{figure}
\centering \epsfig{width=.45\textwidth,file=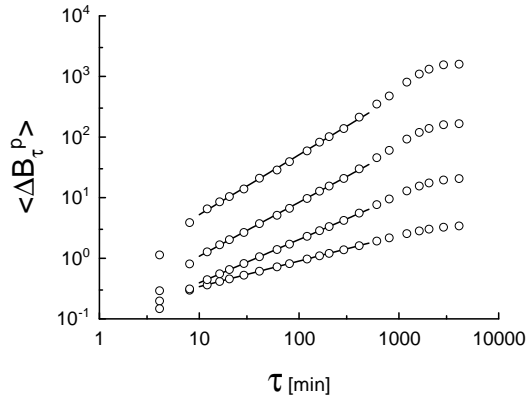}
\vspace{-4cm} \caption{Structure functions of magnetic field {\it
magnitude} in the solar wind plasma as measured by the ACE
magnetometers in nanoTesla for the year 1998 (4 min averages). The
straight lines (the best fits) are drawn to indicate the scaling
law (3) in the inertial range.}
\end{figure}
\begin{figure}
\centering \epsfig{width=.45\textwidth,file=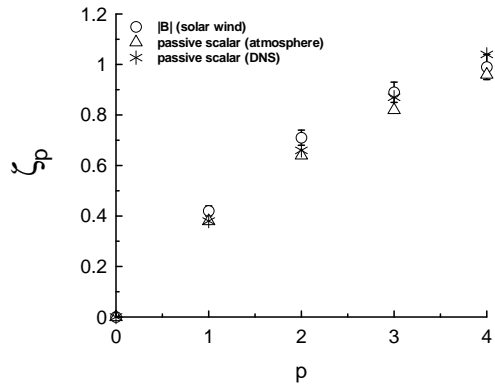}
\vspace{-4cm} \caption{Scaling exponents (3) calculated for $B$ in
the solar wind (circles) and for the passive scalar in the
atmospheric turbulence (triangles, \cite{schmidt}), and in the
direct numerical simulation of 3D fluid turbulence (stars,
\cite{gotoh})}
\end{figure}
\begin{figure}
\centering \epsfig{width=.45\textwidth,file=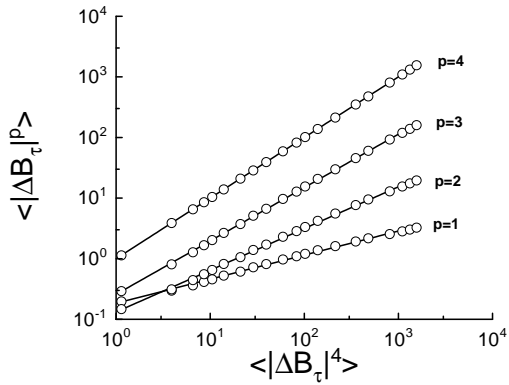}
\vspace{-4cm} \caption{Extended self-similarity (ESS) of the
magnetic field {\it magnitude} in the solar wind plasma. The
straight lines (the best fit) are drawn to indicate the ESS (6).}
\end{figure}

\begin{figure}
\centering \epsfig{width=.45\textwidth,file=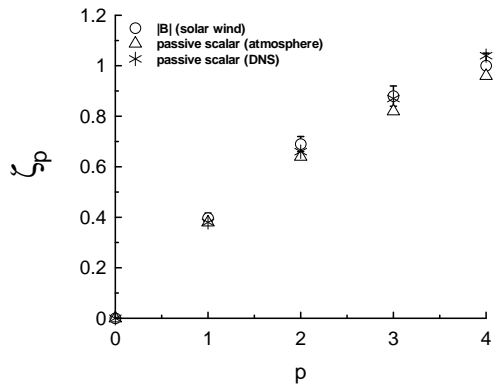}
\vspace{-4cm} \caption{The same as in Fig.\ 3 but using the ESS
method (6) for $B$.}
\end{figure}
Figure 2 shows the scaling of structure functions for the solar
wind data. Slopes of the straight-line fits in the apparently
scaling region provide us the scaling exponents $\zeta_p$; these
are shown in Fig.\ 3 as circles. Triangles in the figure indicate
experimental values obtained for temperature fluctuations in the
atmosphere \cite{schmidt}. The other experimental data
\cite{other1},\cite{other2} are in agreement with each other to
better than 5$\%$. The $\star$ symbols are for the passive scalar
field obtained by numerically solving the advection-diffusion in
three-dimensional turbulence \cite{gotoh}. It is clear that the
exponents for the passive scalar data are in essential agreement
with those for the {\it magnitude} fluctuations of the magnetic
field.

One can, in fact, analyze the solar wind data somewhat differently
using the notion of the extended self-similarity (ESS). Since,
empirically, the fourth order exponent is quite closely equal to 1
for the magnetic field magnitude, i.e.,
$$
\langle |\Delta B_{\tau}|^4\rangle \sim \tau, \eqno{(5)}
$$
we can extend the scaling range (and consequently improve the
confidence with which those exponents are determined) by
redefining them as
$$
\langle |\Delta B_{\tau}|^p \rangle \sim \langle |\Delta
B_{\tau}|^4\rangle^{\zeta_p}. \eqno{(6)}
$$

Figure 4 shows the ESS dependence (6). The slopes of the best-fit
straight lines in this figure provide us with the ESS scaling
exponents $\zeta_p$, which are shown in Fig.\ 5 as circles. The
other symbols have remained unchanged from Fig.\ 3. The shift of
the exponents $\zeta_p$ in comparison to those from ordinary
self-similarity is about 4$\%$, but the scaling interval for ESS
is considerably larger. This increased scaling range is well-known
in other contexts \cite{ben}.

The results shown in Figs.\ 3 and 5 suggest that at least up to
the level of the fourth-order the scaling exponents for the
passive scalars and for the magnitude of the magnetic field are
essentially the same. This is both surprising and
thought-provoking, and needs to be understood further. To this
end, let us return to Eq.\ (1) and specialize \cite{comment}, for
simplicity, to the case of incompressible moving medium ($\nabla
\cdot{\bf v} =0$). Equation (1) can then be rewritten as
$$
\frac{\partial {\bf B}}{\partial t} =- ({\bf v} \cdot \nabla) {\bf
B} + ({\bf B} \cdot \nabla) {\bf v}  +\eta \nabla^2 {\bf B}.
\eqno{(7)}
$$
Let us now consider the equation for the magnitude $B$ of the
magnetic fluctuations given by ${\bf B} = B {\bf n}$, where ${\bf
n}$ is the unit vector with its direction along ${\bf B}$:
$n_i=B_i/B$. Multiplying both sides of Eq.\ (7) by the vector
${\bf n}$ and taking into account that $n_i^2 =1$ we obtain
$$
\frac{\partial B}{\partial t}= -({\bf v}\cdot \nabla) B + \eta
\nabla^2 B + \lambda B, \eqno{(8)}
$$
in which the ``friction-stretching" (or the production)
coefficient $\lambda$ in the last term has the form
$$
\lambda = n_in_j \frac{\partial v_i}{\partial x_j} - \eta \left(
\frac{\partial n_i}{\partial x_j}\right)^2,  \eqno{(9)}
$$
with the indexes $i$ and $j$ representing the space coordinates,
and the summation over repeated indexes is assumed. The first term
on the right hand side of Eq.\ (9) is crucial for any dynamo
effect.

If the statistical behaviors of $\theta$ and $B$ are to be
similar, as suggested by Figs.\ 1, 3 and 5, we should be able to
observe the underlying similarity between Eqs.\ (2) and (8). There
is a major difference corresponding the presence in Eq.\ (8) of
the production term $\lambda B$. However, given the empirical
indications that $B$ and $\theta$ are similar in the inertial
range, it is appropriate to look for circumstances under which the
$\lambda$ term in Eq.\ (8) may be small. The second term in
$\lambda$ is assured to be small because the smallness of the
magnetic diffusivity $\eta$, but difficulties may arise from the
first term on the right hand side of Eq.\ (9).

To eliminate {\it directional} dependencies in Eq.\ (8), let us
make the following conditional average of that equation. That is,
fix the magnitude $B$ in the vector field ${\bf B}=B{\bf n}$ while
performing the average over all realizations of the direction
vector field ${\bf n}$ permitted by the vector equation (7). Let
us denote this ensemble average as $\langle ... \rangle_{{\bf
n}}$. From the definition, this averaging procedure does not
affect $B$ itself, but affects the velocity field ${\bf v}$ and
the ``friction-stretching" coefficient $\lambda$ in Eq.\ (8). We
thus obtain
$$
\frac{\partial B}{\partial t}= -(\langle {\bf v} \rangle_{{\bf n}}
\cdot \nabla) B + \eta \nabla^2 B + \langle \lambda \rangle_{{\bf
n}}B. \eqno{(10)}
$$
It is worth emphasizing that the solutions of the original
equation (7) satisfy Eqs.\ (8) and (10), but not all possible
formal solutions of the Eqs.\ (8) and (10) satisfy Eq.\ (7);
similarly, not all formal solutions of Eq.\ (10) satisfy Eq.\ (8)
while all solutions of Eq.\ (8) do satisfy Eq.\ (10). Restricting
comments to the relationship between Eqs.\ (8) and (10), the
solutions of the two equations are the same only if the initial
conditions are the same and if realizations of $\langle {\bf v}
\rangle_{\bf n}$ and of $\langle \lambda \rangle_{\bf n}$, related
to these initial conditions by the conditional average procedure,
are obtained from solutions applicable to Eq.\ (8).

Returning now to Eq.\ (10), the conditionally averaged velocity
field $\langle {\bf v} \rangle_{{\bf n}}$ may posses statistical
properties that are different from those of the original velocity
field ${\bf v}$, and there can be circumstances under which
$\langle \lambda \rangle_{{\bf n}}=0$, or small. If so, the
similarity between Eqs.\ (2) and (10) (and, consequently, Eq.\
(8)) can be the basis for the similarity in statistical properties
of their solutions. Therefore, finding conditions under which
$\langle \lambda \rangle_{{\bf n}}=0$, or small, seems to be a
useful exercise.

It is, however, difficult to guess {\it a~priori} when $\langle
\lambda \rangle_{{\bf n}}$ is negligible, because there is no
small parameter for the stretching part of $\lambda$. Therefore,
let us consider a generic set of conditions, presumably for the
inertial range, which can result in $\langle n_in_j
\partial v_i/\partial x_j \rangle_{{\bf n}} =0$. This can be a
combination of isotropy, which yields
$$
\langle n_in_j\rangle_{{\bf n}} =0 ~~~~(i\neq j)
$$
and
$$
 \langle n_1^2\rangle_{{\bf n}}=\langle n_2^2\rangle_{{\bf n}}=
\langle n_3^2 \rangle_{{\bf n}}, \eqno{(11)}
$$
and statistical independence
$$
\langle n_in_j \varphi \rangle_{{\bf n}} = \langle n_in_j
\rangle_{{\bf n}}\langle  \varphi \rangle_{{\bf n}}, \eqno{(12)}
$$
where $\varphi = \partial v_k/\partial x_l$ for arbitrary $k$ and
$l$.

We should emphasize that the conditional average indicated by
$\langle \dots \rangle_{{\bf n}}$ and the global average indicated
by $\langle \dots \rangle$ are quite different; because of this,
the quantity $B$ in (10) remains a fluctuating variable. To
eliminate the stretching part from the conditionally averaged
coefficient $\langle \lambda \rangle_{{\bf n}}$---this being
critical for explaining the observed similarity in the scaling of
structure functions between $B$ and $\theta$---one does not need
to satisfy conditions (11) and (12) for all realizations of the
magnetic field ${\bf B}$, but only for the subset of realizations
that gives the main statistical contribution to the structure
functions (3). Let us name this subset of realizations as {\it I}.
The structure functions (3) depend on the statistical properties
of the {\it increments} with respect to $\tau$, namely $\Delta
B_{\tau}$, belonging to the inertial range of scales. One of the
consequences of intermittency is that the statistical properties
of the increments are essentially different from those of the
field ${\bf B}$ itself. Therefore, the subset {\it I} need not
generally coincide with the subset {\it G}, say, that gives the
main statistical contribution to the {\it global} average $\langle
n_in_j \partial v_k/\partial x_l \rangle$. This means, in
particular, that the conditions (11) and (12) can be valid for the
inertial interval (i.e. for subset {\it I}), while globally (i.e.
for subset {\it G}) these conditions could well be violated.

We now use conditions (11) and (12) in the presence of the
incompressibility condition $\partial v_i/\partial x_i=0$ and
obtain
$$
\langle \lambda \rangle_{{\bf n}} = -\eta \langle \left(
\frac{\partial n_i}{\partial x_j}\right)^2 \rangle_{{\bf n}}.
\eqno{(13)}
$$
That is, the difference between the passive scalar equation (2)
and the conditionally averaged equation (10) for $B$ is reduced to
pure ``friction" with the friction coefficient given by (13).
Equation (10) can then be reduced in Lagrangian variables to
$$
\frac{dB}{dt}=\langle \lambda \rangle_{{\bf n}} B, \eqno{(14)}
$$
with the ``multiplicative noise" $\langle \lambda \rangle_{{\bf
n}}$ given by Eq.\ (13). Weak diffusion of Lagrangian ``particles"
can be described as their wandering around the deterministic
trajectories. Introduction of a weak diffusion is equivalent to
introduction of additional averaging in Eq.\ (14) over random
trajectories \cite{zeld1}. The small parameter $\eta$ in (13) and
(14) will then determine a slow time in comparison with the time
scales in the inertial interval and will therefore not affect
scaling properties of $B$ in the inertial interval. This explains
the similarity of scaling between $B$ and $\theta$.

\section{The magnetic field gradients}

If one can neglect the last term in the right-had side of the equation 
(10) in the inertial range, then one can readily derive equation for 
the magnitude gradients ${\bf G} \equiv \nabla B$
$$
\frac{\partial G_i}{\partial t}=-\langle v_j \rangle_{{\bf n}}\frac{\partial G_i}{\partial
x_j} - \frac{\partial \langle  v_j \rangle_{{\bf n}}}{\partial x_i}G_j + \eta \frac{\partial^2
G_i}{\partial x_j^2},  \eqno{(15)}
$$
The magnitude $G$ of the gradient is determined by
${\bf G} = G {\bf g}$, where ${\bf g}$ is the unit vector with its
direction along vector ${\bf G}$. Multiplying both sides of Eq.\
(15) by $g_i$, making summation over $i$, and taking into account
of the fact that $g_i^2 =1$, we obtain
$$
\frac{\partial G}{\partial t}= -(\langle {\bf v} \rangle_{{\bf n}}\cdot \nabla) G + \eta
\nabla^2 G - \lambda' G, \eqno{(16)}
$$
which is formally similar to Eq.\ (2) (cf. Eq.\ (8)) except for the last term in
(16). The coefficient $\lambda'$ in this term has the form (cf. Eq.\ (9))
$$
\lambda' = g_ig_j \frac{\partial \langle  v_i \rangle_{{\bf n}}}{\partial x_j} 
+ \eta \left(\frac{\partial g_i}{\partial x_j}\right)^2.  \eqno{(17)}
$$

One can see remarkable similarity to the situation described in previous section. 
This similarity suggests applying ensemble conditional average 
similar to that described above. Fix the magnitude $G$ in the vector 
field ${\bf G} = G{\bf n}$ while performing the average over all realizations of
the direction vector field ${\bf g}$ permitted by equation (15).
Let us denote this ensemble average as $\langle ... \rangle_{{\bf
g}}$. From the definition, this averaging procedure does not
affect $G$ itself, but modifies the velocity field 
$\langle {\bf v} \rangle_{{\bf n}}$,
which in turn modifies the coefficient $\lambda'$ in Eq.\ (16). We
may write 
$$
\frac{\partial G}{\partial t}= -(\langle \langle {\bf v} 
\rangle_{{\bf n}}\rangle_{{\bf g}}
\cdot \nabla) G + \eta \nabla^2 G + \langle \lambda' \rangle_{{\bf
g}}G. \eqno{(18)}
$$

The further analysis can be performed precisely as it has been
done above for magnitude of magnetic filed itself. Two main consequences
of the directional average: isotropization and smoothing of the
velocity field $\langle {\bf v} \rangle_{{\bf n}}$ and the 
"nullification" of the production term in the
conditionally averaged equations, can have different
proportions in these two cases. For the above considered case with
magnitude of magnetic field the conditionally averaged (on the
directions of the magnetic field) velocity was still enough
strongly fluctuating to belong to the inertial range paradigm. It
is possible, however, that the conditional average on the
directions of the magnitude {\it gradients} can smooth the already smoothed
velocity field $\langle {\bf v} \rangle_{{\bf n}}$ to a substantially 
non-fluctuating state. In the last case the essential point is that 
the twice conditionally averaged velocity $\langle \langle {\bf v} 
\rangle_{{\bf n}} \rangle_{{\bf g}}$ is smoothed
substantially in comparison with ${\bf v}$, while the fluctuation
of $G$ itself is still rapid in the diffusion-advection equation
(18) (because it remains in tact under the conditional average, by
virtue of its definition). Under these typical circumstances, the
natural expectation (see, for instance, \cite{ynaog} and
references therein) is that the space autocorrelation function 
can be characterized by a logarithmic behavior \cite{chertkov} 
given by
$$
C(r)= \frac{\langle G(r) G(0) \rangle}{\langle G(0)^2 \rangle}
\sim \ln \left(\frac{L}{r} \right), \eqno{(19)}
$$
The result owes itself to the pioneering work of Batchelor \cite{batc},
\cite{Batchelor2} who applied this general idea to the viscous-convection 
range of passive scalar fluctuations with large $Pr$. While the two
contexts are quite different, they are the same in the sense that
the velocity field is smooth whereas the advected quantity is
strongly fluctuating.

For turbulent flows the Taylor hypothesis is generally used to
interpret the data. This hypothesis states that the intrinsic 
time dependence of the magnetic field can be ignored when the 
turbulence is convected past the probes at nearly constant speed. 
With this hypothesis, the temporal dynamics should reflect the 
spatial one \cite{b}. Equation (19) can be rewritten in the temporal 
terms ($r \rightarrow \tau$ $L \rightarrow \tau_0$) 
$$
C(\tau )= \frac{\langle G(\tau) G(0) \rangle}{\langle G(0)^2
\rangle} \sim \ln \left(\frac{\tau_0}{\tau} \right). \eqno{(20)}
$$

This is seen from Figs. 6 and 7 (in the semi-log scales) to apply quite
precisely for the solar wind data (the solid
straight line indicates the logarithmic dependence (20)) obtained 
from Advanced Composition Explorer (ACE) satellite magnetometers 
for the year 1998 (Fig. 6) and for the year 2004 (Fig. 7). 

The so-called pumping scale \cite{chertkov} $L$ (or $\tau_0$)
in (19),(20) can be defined from the Figs. 6 and 7 as intersection of the
solid straight line with the temporal $\tau$ co-ordinate axis: $\tau_0
\simeq 24h \pm 2h$.

\begin{figure}\vspace{-2cm}
\centering \epsfig{width=.45\textwidth,file=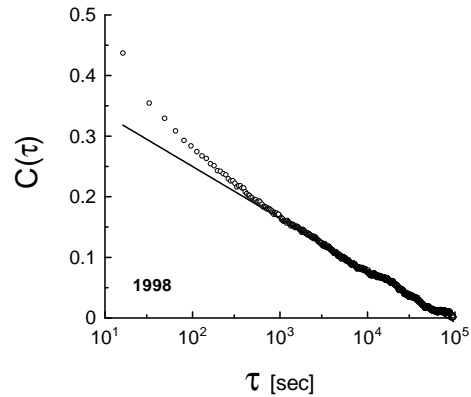}
\vspace{-4.5cm} \caption{Autocorrelation function $C(\tau)$ of the
magnitude of the gradient of $B$ plotted against $\log \tau$, for
the ACE-1998 solar wind data. }
\end{figure}
\begin{figure}\vspace{-0.4cm}
\centering \epsfig{width=.45\textwidth,file=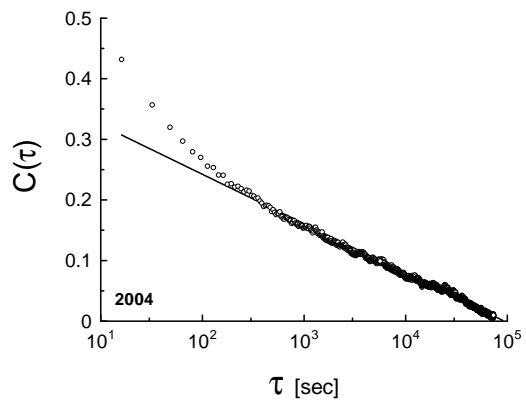}
\vspace{-4.5cm} \caption{The same as in Fig. 6 but for 2004 year.}
\end{figure}

\section{Geophysical consequences}
In order to get away from the effects of the Earth's magnetic field, 
the ACE spacecraft orbits at the L1 libration point which is a point of 
Earth-Sun gravitational equilibrium about 1.5 million km from Earth and 
148.5 million km from the Sun. Let us recall that the Earth's magnetosphere 
extends into the vacuum of space from approximately 80 to 60,000 kilometers 
on the side toward the Sun. Therefore, the observed in Figs. 6,7 approximately 
24h pumping time-scale of the magnetic field gradients cannot be 
a consequence of the 24h-period of the Earth's rotation. The pumping scale 
certainly corresponds to a process in the solar wind plasma itself. It could be 
a characteristic scale of the phase coherent structures in the solar wind, 
which can be distinguished from incoherent fluctuations in this case \cite{rg2} 
(see more about characteristic periods in solar wind plasma in the recent review 
\cite{zmd}). It is now believed that energy of these coherent (pumping) 
large-scale structures is released via shear instabilities in the solar wind 
plasma. 

It is possible that for the very long time of the coexistence of Earth and 
of the solar wind the weak interaction between the solar wind and Earth could 
lead to stochastic synchronization between the Earth's rotation and the 
characteristic time-scale (the pumping scale) of the solar wind magnetic field 
gradients. This synchronization could transform an original period of the Earth's 
rotation to the period close to the pumping scale, which we observe at present time. \\

I thank ACE/MAG instrument team as well as the ACE
Science Center for providing the data and support, T. Gotoh for
providing Ref. \cite{gotoh} before publication. I also thank K.R. Sreenivasan for inspiring 
cooperation, and D. Donzis, J. Schumacher, and V. Steinberg for comments and suggestions.


\begin{thebibliography}{99}
\bibitem{barn1} Barnes A., 1979, Hydrodynamic Waves and Turbulence
in the solar Wind, in Solar System Plasma Physics, Vol. I, eds.
Parker E.N. Kennel C.F., and Lanzerotti L.J., p. 251,
North-Holland.
\bibitem{barn2} Barnes A., 1981, J. Geophy. Res. {\bf 86}, 7498.
\bibitem{bel} Belcher J.W. and Davis L. Jr., 1971, J. Geophys.
Res., {\bf 76}, 3534.
\bibitem{b} Bershadskii A., 2003, Phys. Rev. Lett., {\bf 90}, 041101.
\bibitem{bs} Bershadskii A. and Sreenivasan K.R., 2004, Phys. Rev. Lett., {\bf 93},
064501.
\bibitem{biskamp1} Biskamp D., 1993, Nonlinear Magnetohydrodynamics
(Cambridge Univ. Press, Cambridge, UK.).
\bibitem{bur1} Burlaga L.F. and Ness N.F., 1998, J. Geophys. Res.,
{ \bf 103}, 29719.
\bibitem{bur2} Burlaga L.F., Wang C., Richardson J.D., and Ness
N.F., 2003, Ap. J., {\bf 585}, 1158.
\bibitem{gold1} Goldstein M.L., Burlaga L.F. and Matthaeus, 1984,
J . Geophys. Res., {\bf 89}, 3747.
\bibitem{gold2} Goldstein M.L., 2001, Astrophys.\ Space Sci.\ {\bf 227}, 349.
\bibitem{hartlep} Hartlep T., Matthaeus W.H., Padhye N.S., and
Smith C.W., (2000), {\bf 105}, 5135.
\bibitem{o} Oughton S., Priest E.R. and Matthaeus W.N., 1994, J. Fluid Mech.,
{\bf 280}, 95.
\bibitem{rg2} Roberts D.A. and Goldstein M.L., 1987, J. Geophys. Res.,
A, {\bf 92}, 10105.
\bibitem{zmd} Zhou Ye, Matthaeus W.H. and Dmitruk P., 2004, Rev. Mod.
Phys., {\bf 74}, 1015.
\bibitem{shr} B.I. Shraiman and E.D. Siggia, Nature {\bf 405}, 639 (2000).
\bibitem{fal} G. Falkovich, K. Gawedzki and M. Vergassola, Rev.\ Mod.\
Phys.\ {\bf 73}, 913 (2001).
\bibitem{cohen} Y. Cohen, T. Gilbert
and I. Procaccia, Phys.\ Rev.\ E {\bf 65}, 026314 (2002).
\bibitem{ching} E.S.C. Ching, Y. Cohen, T. Gilbert and I.
Procaccia, Phys.\ Rev.\ E {\bf 67}, 016304 (2003).
\bibitem{bur} L. F. Burlaga, Interplanetary Magnetohydrodynamics (Oxford University
Press, New York) 1995.
\bibitem{gold}
M.L. Goldstein, Astrophys.\ Space Sci.\ {\bf 227}, 349 (2001).
\bibitem{gfn} Gotoh T., Fukayama D. \& Nakano T., 2002, Phys. Fluids, {\bf 14},
1065.
\bibitem{my} A.S. Monin and A.M.
Yaglom, Statistical Fluid Mechanics: Mechanics of Turbulence,
vol.\ 2 (MIT Press, Cambridge) 1975.
\bibitem{schmidt} F. Schmitt, D. Schertzer, S. Lovejoy and Y. Brunet, Europhys.\ Lett.\ {\bf 34}, 195 (1996).
\bibitem{gotoh} T. Watanabe and T. Gotoh, Statistics of Passive Scalar in Homogeneous
Turbulence (submitted).
\bibitem{other1} R.A. Antonia, E. Hopfinger, Y. Gagne and F. Anselmet,
Phys.\ Rev.\ A {\bf 30}, 2705 (1984).
\bibitem{other2} C. Meneveau, K.R. Sreenivasan, P. Kailasnath and
M.S. Fan, Phys.\ Rev.\ A {\bf 41}, 894 (1990).
\bibitem{ben} R.
Benzi, L. Biferale, S. Ciliberto, M.V. Struglia and R.
Tripiccione, Physica D {\bf 96}, 162 (1996).
\bibitem{comment} This is not overly restrictive in the inertial range; see, e.g., M.L.
Goldstein, D.A. Roberts and W.H. Matthaeus, Annu. Rev. Astron.
Astrophys. {\bf 33}, 283 (1995).
\bibitem{zeld1} Ya.B. Zeldovich, A.A. Ruzmaikin and D.D. Sokoloff,
Magnetic Fields in Astrophysics (Gordon and Breach) 1983; Ya.B.
Zeldovich, B. Molchanov, A.A. Ruzmaikin, D.D. Sokolov  Sov. Phys.\
Usp.\ {\bf 30}, 353 (1987).
\bibitem{sbn2} Sreenivasan K.R., Bershadskii A., and Niemela J.J.,
2005, Phys. Rev. E {\bf 71}, 035302(R).
\bibitem{ynaog} Yuan G.-C., Nam K., Antonsen T.M.. Ott E., and Guzdar P.N.,
2000, Chaos {\bf 10}, 39.
\bibitem{chertkov} Chertkov M., Falkovich G., Kolokolov I., and
Lebedev V., 1995, Phys. Rev. E {\bf 51}, 5609.
\bibitem{batc} Batchelor G.K. , 1959, J.\ Fluid Mech. {\bf 5}, 113.
\bibitem{Batchelor2} Batchelor, G.K., 1970, An
Introduction to Fluid Dynamics, Cambridge University Press,
Cambridge.




\end{thebibliography}
\end{document}